\documentstyle[aps,prl,multicol,epsf]{revtex}

\begin{document}

\title{
Geometrical frustration induced (semi-)metal to insulator transition 
}

\author{Satoshi Fujimoto}
\address{
Department of Physics,
Kyoto University, Kyoto 606-8502, Japan
}

\date{\today}
\maketitle
\begin{abstract}
We study the low-energy properties of the geometrically frustrated
Hubbard model on a three-dimensional pyrochlore lattice 
and a two-dimensional checkerboard lattice
on the basis of the renormalization group method 
and mean field analysis.
It is found that in the half-filling case, 
a (semi-)metal to insulator transition (MIT) occurs.
Also, in the insulating phase, which has a spin gap, the spin rotational
symmetry is not broken, while charge ordering exists.
The results are applied to the description of 
the MIT observed in the pyrochlore system 
${\rm Tl_2Ru_2O_7}$.
\end{abstract}

\pacs{PACS numbers: }

\begin{multicols}{2}

Recently, the role of geometrical frustration in both localized and
itinerant electron systems has attracted 
renewed interest
\cite{rami,pyroex,pyro,check,kagome,fuji,take,mand}.
In localized systems, the possibility of exotic phase such as a spin liquid
without long-range magnetic order 
has been extensively investigated
both experimentally\cite{rami,pyroex,take,mand} and 
theoretically\cite{pyro,check,kagome,fuji}.
In itinerant systems, 
the presence of charge degrees of
freedom provides a route for the relaxation of magnetic frustration.
However, when electron correlation is sufficiently strong,  
the magnetic frustration may still affect the low-energy properties 
significantly.
For example, it has been found experimentally that 
some pyrochlore oxides, such as ${\rm Tl_2Ru_2O_7}$ and ${\rm Cd_2Os_2O_7}$,
exhibit a (semi-)metal to insulator transition (MIT) at finite critical
temperatures\cite{take,mand}. 
Since such systems possess the fully-frustrated lattice structure, 
referred to as a network of corner-sharing tetrahedra (that is, a pyrochlore
lattice), the magnetic properties of 
the insulating phase are yet largely unexplained.
Moreover the mechanism of the MITs observed in these systems 
is still an open problem.
Geometrical frustration may play an important role in the MITs.
From this point of view, in the present paper, 
we study the interplay between electron correlation
and geometrical frustration in the Hubbard model 
on a three-dimensional (3D) pyrochlore lattice
and on a two-dimensional (2D) checkerboard lattice,
the so-called 2D pyrochlore (FIG.1).
Although real pyrochlore oxides have electronic structure
composed of $t_{2g}$ orbitals, the present study on these simpler
single-band models
may provide important insight into the role of geometrical frustration
in MIT. 
Furthermore, ${\rm Tl_2Ru_2O_7}$ has, apart from the $t_{2g}$ band, 
a nearly half-filled Tl $s$ band, 
whose important features are described 
by the 3D pyrochlore Hubbard model\cite{ogu}.
We believe that this model may provide a useful understanding of 
the MIT undergone in this material.


The non-interacting energy bands of these two Hubbard models 
have a common interesting feature: 
They consist of a flat band (or two degenerate flat bands) 
on the upper band edge and
a dispersive band that is tangent to the flat band 
(or flat bands) at the $\Gamma$ point\cite{mielke}.
According to graph theory, this band structure stems from
the geometrical frustration inherent in the pyrochlore lattice.
In the half-filling case, $n=1$, on which we concentrate henceforth,
the dispersive band is fully occupied, 
and the flat band(s) is empty.
In the non-interacting half-filling case,
the system is in a semi-metal state,
since the Fermi velocity is vanishing, though there is no excitation gap. 
We study how this state is affected by electron correlation.

\begin{figure}[h]
\centerline{\epsfxsize=7.5cm \epsfbox{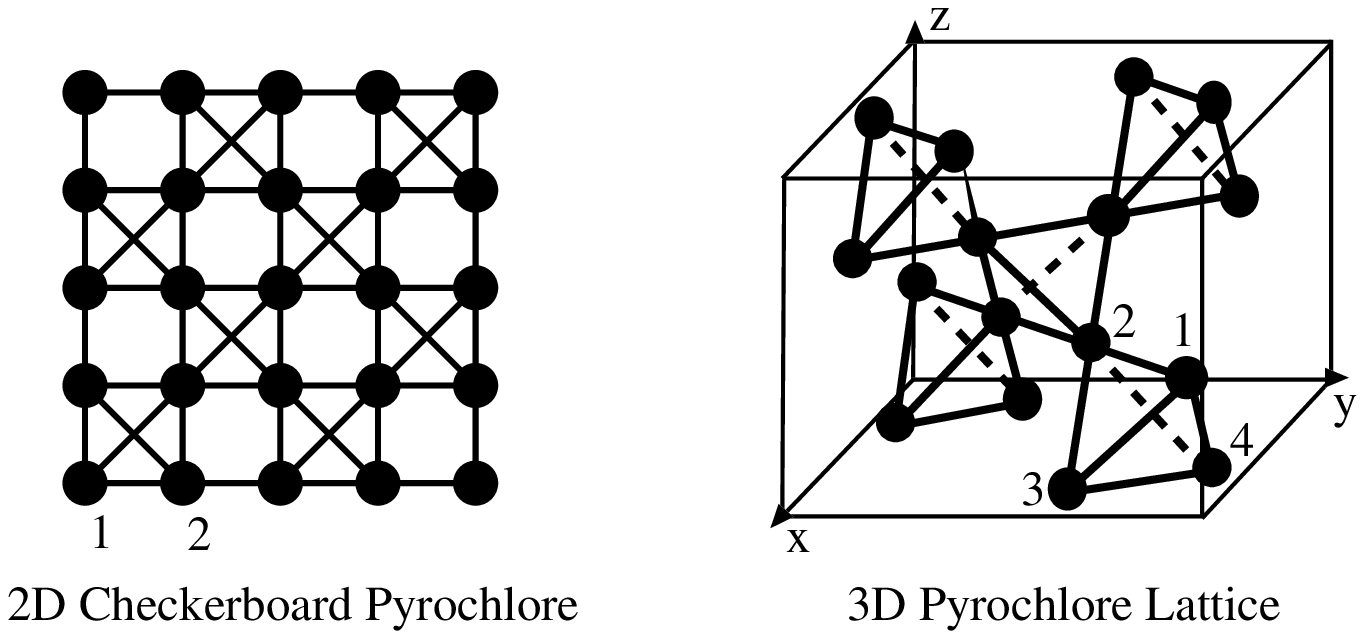}}
{FIG. 1. 2D and 3D pyrochlore lattices.}
\end{figure}

Diagonalizing the kinetic term, we write the Hamiltonian as
\begin{eqnarray}
&&H=\sum_{\mu=1}^{m}\sum_{k\sigma}E_{k\mu}a^{\dagger}_{k\mu\sigma}
a_{k\mu\sigma} \nonumber \\
&&\qquad +\frac{U}{N}\sum_{k,k',q}\sum_{\alpha\beta\gamma\delta}
\Gamma^0_{\alpha\beta\gamma\delta}(k-q,k'+q;k',k) \nonumber \\
&&\qquad\times a^{\dagger}_{k-q\alpha\uparrow}
a^{\dagger}_{k'+q\beta\downarrow}
a_{k'\gamma\downarrow}a_{k\delta\uparrow}, \label{hamil} \\
&&\Gamma^0_{\alpha\beta\gamma\delta}(k_1,k_2;k_3,k_4)= 
\sum_{\nu=1}^m 
s_{\nu\alpha}(\mbox{\boldmath $k_1$})
s_{\nu\beta}(\mbox{\boldmath $k_2$})
s_{\nu\gamma}(\mbox{\boldmath $k_3$})
s_{\nu\delta}(\mbox{\boldmath $k_4$}), \nonumber 
\end{eqnarray}
where $m=2$ in the 2D case, and 
$m=4$ in the 3D case.
In the 2D case, $E_{k1}=2$, $E_{k2}=4\cos k_x \cos k_y-2$, 
$s_{11}(\mbox{\boldmath $k$})=s_{22}(\mbox{\boldmath $k$})
=\sin(\frac{k_x+k_y}{2})/\sqrt{1-\cos k_x\cos k_y}$, and 
$s_{21}(\mbox{\boldmath $k$})=-s_{12}(\mbox{\boldmath $k$})
=\sin(\frac{k_x-k_y}{2})/\sqrt{1-\cos k_x\cos k_y}$.
In the 3D case, $E_{k1}=E_{k2}=2$, $E_{k3}=-2+2\sqrt{1+t_k}$, and  
$E_{k4}=-2-2\sqrt{1+t_k}$, 
with $t_k=\cos(2k_x)\cos(2k_y)+\cos(2k_y)\cos(2k_z)+\cos(2k_z)\cos(2k_x)$.
The form of $s_{\mu\nu}(\mbox{\boldmath $k$})$ 
in the 3D case is given in Ref\cite{fuji}.
The annihilation operator of electrons at the $\mu$-th site
in a unit cell is given by 
$c_{k\mu\sigma}=\sum_{\nu=1}^ms_{\mu\nu}(\mbox{\boldmath $k$})a_{k\nu\sigma}$.

As shown in Ref\cite{fuji}, 
the perturbative calculation in $U$ for the above Hamiltonian 
suffers from divergences at third and higher order, 
due to the presence of the flat band(s).
To treat the divergences in a controlled manner, we apply
the renormalization group (RG) method.
In previous studies of electron systems\cite{frg}, 
a momentum cutoff that separates 
the neighborhood of the Fermi surface from the higher momentum part
is introduced.
However, in the presence of the flat band(s), this procedure is not applicable.
To overcome this problem, 
we introduce the infrared energy cutoff $\Lambda$ in the following manner:
$\psi_{\mu\sigma}(k,\varepsilon_n)
=\psi^{>}_{\mu\sigma}(k,\varepsilon_n)\Theta(|\varepsilon_n|-\Lambda)
+\psi^{<}_{\mu\sigma}(k,\varepsilon_n)\Theta(\Lambda-|\varepsilon_n|)$.
Here, $\psi_{\mu\sigma}(k,\varepsilon)$ 
is the Grassmann field corresponding to $a_{k\mu\sigma}$.

Using a standard method, we can obtain
the RG equations of the single-particle
self-energy, $\Sigma_{\mu\nu}(k)$, and the 4-point vertex functions, 
$\Gamma_{\alpha\beta\gamma\delta}(k_1,k_2;k_3,k_4)$ up to the one-loop order.
In our systems, there are six species of 4-point vertices, 
as shown in FIG.2(a), apart from the spin degrees
of freedom and the two-fold degeneracy of the flat bands in the 3D case. 
We assume that the momentum dependences of the 4-point vertex functions are
given mainly by $\Gamma^0(k_1,k_2;k_3,k_4)$ 
in the renormalization processes. 
This is made explicitly by replacing $\Gamma_{abab}(k_1,k_2;k_3,k_4)$ with
$g_1\Gamma^0_{abab}(k_1,k_2;k_3,k_4)$, $\Gamma_{bbba}(k_1,k_2;k_3,k_4)$
with $g_4\Gamma^0_{bbba}(k_1,k_2;k_3,k_4)$, etc.
This approximation is fairly good, because 
in the vicinity of the $\Gamma$ point,
where the most important scattering processes occur,
the band structure is almost isotropic.
Because the flat bands are empty and the dispersive band
is fully occupied, the particle-particle processes between the flat bands 
and the particle-hole processes
between the flat bands and the dispersive band give the leading
singular contributions (see FIG.2(b)). 
We take into account these contributions in the derivation
of the RG equations.
We found from the analysis of the RG equations that,
among the six running couplings, $g_2$, $g_3$, and $g_5$ are
irrelevant in the low energy limit.
The RG equations for the other couplings, $g_1$, $g_4$, and $g_6$ 
are written
\begin{eqnarray}
\frac{d g_{1s}}{d l}&=&
-\frac{ag_{4s}^2e^l}{\Lambda_0}+\frac{b(\Lambda_0e^{-l})^{\eta}}{4}
(g_{1s}^2+6g_{1s}g_{1t}-3g_{1t}^2), 
\label {rge1}\\
\frac{d g_{1t}}{d l}&=&
\frac{b(\Lambda_0e^{-l})^{\eta}}{4}
(g_{1s}^2-2g_{1s}g_{1t}+5g_{1t}^2), 
\label{rge2}\\
\frac{d g_{4s}}{d l}&=&
-\frac{ag_{4s}g_{6s}}{\Lambda_0}e^l
+\frac{b(\Lambda_0e^{-l})^{\eta}}{4}
(g_{1s}g_{4s}+3g_{4s}g_{1t}), \label{rge3}\\
\frac{d g_{6s}}{d l}&=&
-\frac{ag_{6s}^2}{\Lambda_0}e^l, \label{rge4}
\end{eqnarray}
where $\eta=(d-2)/2$, 
$l={\rm ln}(\Lambda_0/\Lambda)$, with $\Lambda_0$ the band width, 
and $d$ is the spatial dimension.
The couplings $g_{i s}$ and $g_{i t}$ 
denote the spin singlet and triplet parts, 
respectively.
In the 2D case, $a=\sum_k(s_{11}^4(k)-s_{11}^2(k)s_{12}^2(k))/2
$, $b=1/(32t)$, and in the 3D case, $a=\sum_k[(s_{11}^2(k)+s_{12}^2(k))^2
-(s_{11}(k)s_{21}(k)+s_{12}(k)s_{22}(k))^2]/2$, and $b\approx 0.0775/t^{3/2}$.
In the derivation of these equations for the 3D case, 
we have used the fact that in the vicinity of the $\Gamma$ point,
the two degenerate flat bands 
do not mix with each other in scattering processes. 
Thus in this case the two-fold degeneracy just gives 
an overall factor of $2$.

\begin{figure}[h]
\centerline{\epsfxsize=7.2cm \epsfbox{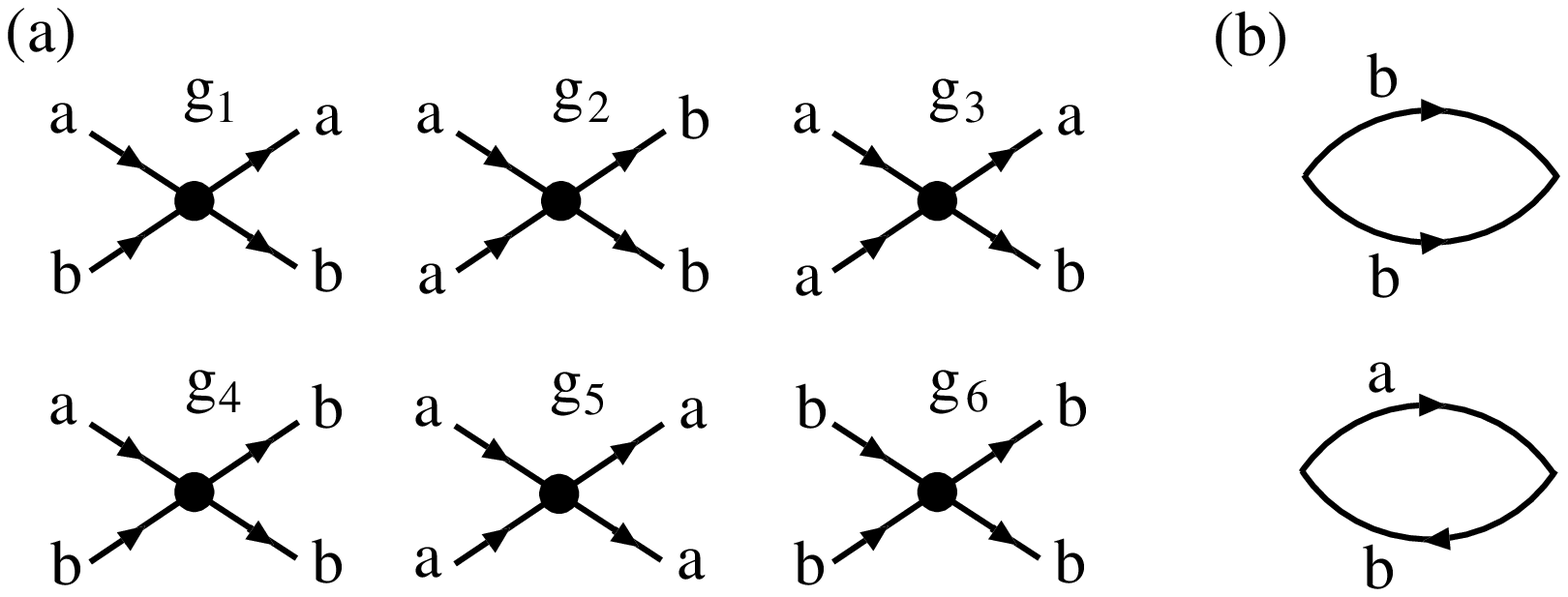}}
{FIG. 2(a) The six species of 4-point vertices. Here
``$a$'' and ``$b$'' indicate 
the dispersive band and the flat band, respectively. 
(b) The leading singular bubble diagrams.}
\end{figure}

{\it 2D checkerboard pyrochlore--}
We first consider the 2D case, whose theoretical treatment is simpler.
We solved the RG equations (\ref{rge1})-(\ref{rge4}) numerically
for a particular set of parameter values, and
obtained the RG flow shown in FIG.3(a). 
We found that for any small value of $U/t$, 
$g_{1t}$ flows into the strong-coupling regime. This indicates some
instability in this channel.
Although $g_{1s}$ also scales into the strong-coupling regime, 
it is sub-dominant compared to $g_{1t}$.
We also show in FIG.3(a) the RG flows of the couplings $3g_{1t}-g_{1s}$ 
and $g_{1s}+g_{1t}$, which are related to the charge and spin
susceptibilities, respectively.
We see that some instability appears in the charge degrees of freedom.
To elucidate the nature of this instability, 
we write the RG equations for the single-particle self-energy
in the following form:
\begin{eqnarray}
&&\frac{d(\Sigma_{12\uparrow\uparrow}^{\Lambda}
+\Sigma_{12\downarrow\downarrow}^{\Lambda})}{d l}
=2b(3g_{1t}-g_{1s})
(\Sigma_{12\uparrow\uparrow}^{\Lambda}
+\Sigma_{12\downarrow\downarrow}^{\Lambda}) \label{selfrg} \\
&&\frac{d(\Sigma_{12\uparrow\uparrow}^{\Lambda}
-\Sigma_{12\downarrow\downarrow}^{\Lambda})}{d l}
=2b(g_{1s}+g_{1t})
(\Sigma_{12\uparrow\uparrow}^{\Lambda}
-\Sigma_{12\downarrow\downarrow}^{\Lambda}) \\
&&\frac{d\Sigma_{12\uparrow\downarrow}^{\Lambda}}{d l}
=2b(g_{1s}+g_{1t})\Sigma_{12\uparrow\downarrow}^{\Lambda}.
\end{eqnarray}
In the derivation of these equations, we have ignored
the diagonal self-energy, which are not important in the following argument, 
and expanded the RG equations 
up to the first order in $\Sigma^{\Lambda}_{12}$.
Because the strongest divergence of the 4-point vertex appears 
in $3g_{1t}-g_{1s}$ (see FIG.3(a)), the off-diagonal
self-energy $\sum_{\sigma}\Sigma_{12\sigma\sigma}$
becomes non-zero at some critical $\Lambda=\Lambda_c$.
This is easily seen by solving (\ref{selfrg}), which gives
$\sum_{\sigma}\Sigma^{\Lambda}_{12\sigma\sigma}=
\sum_{\sigma}\Sigma^{\Lambda_0}_{12\sigma\sigma}
{\rm exp}[2b\int^l_0 dl'(3g_{1t}-g_{1s})]$.
Note that $\Sigma_{12\sigma\sigma}^{\Lambda_0}$ is vanishing
in the vicinity of the $\Gamma$ point, because of the momentum dependence
of $s_{\mu\nu}(k)$. Thus for $\Lambda=\Lambda_c$ at which value
$3g_{1t}-g_{1s}$ is divergent, 
$\sum_{\sigma}\Sigma_{12\sigma\sigma}^{\Lambda_c}$ becomes non-zero.

The above RG analysis implies the existence of a mean field solution
for which the order parameter is given by $\Delta_k\equiv
\sum_{\sigma}\Sigma_{12\sigma\sigma}(k)
\sim \sum_{\sigma={\uparrow\downarrow}}
\langle a_{k 1\sigma}^{\dagger}a_{k 2\sigma}\rangle$.
This state is characterized by electron-hole pairing with parallel spins, 
which leads to the formation of both spin and charge gaps
preserving the spin rotational symmetry. 
According to the numerical analysis of the RG equations
(\ref{rge1})-(\ref{rge4}),
$g_{4s}$ is mainly renormalized by the first term
of the right-hand side of (\ref{rge3}).
Then, the renormalized coupling $g_{4s}$ is 
approximately given by RPA-like expressions.
As a result,  
we obtain the self-consistent gap equation for $\Delta_k$
expressed diagramatically in FIG.4.
The transition temperature is determined from the linearized gap equation, 
\begin{eqnarray}
\Delta_k&=&\sum_{q,k'}\Pi(k,q-k)
G_{11}(q-k)G_{22}(q-k) \nonumber \\
&\times&\Pi(k',q-k)G_{11}(k')G_{22}(k')\Delta_{k'}, \label{gap}
\end{eqnarray}
where $G_{\mu\mu}(k)=1/(\varepsilon_n-E_{k\mu})$,  
$k=({\rm i}\varepsilon,\mbox{\boldmath $k$})$, and
$\Pi(k,k')=\sum_{\nu=\pm}\nu
Ut^{\nu}(\mbox{\boldmath $k$},\mbox{\boldmath $k'$)}
/(1-c_{\nu}D(k+k'))$,  
$t^{\pm}(\mbox{\boldmath $k$},\mbox{\boldmath $k'$})
=(s_{11}(\mbox{\boldmath $k$})s_{12}(\mbox{\boldmath $k'$})
\pm s_{12}(\mbox{\boldmath $k$})s_{11}(\mbox{\boldmath $k'$}))^2/2$
with $D(q)=-TU\sum_{n,k} G_{11}(k)G_{11}(q-k)$,
$c_{+}=2a$, and $c_{-}=\sum_k s_{11}^2$.
Here we have ignored the diagonal self-energy. 
Equation (\ref{gap}) implies that the gap function can be written
$\Delta_k=s_{11}(k)s_{12}(k)\Delta_0$, 
where $\Delta_0$ is a constant. From (\ref{gap}), we have
$\Delta_0=\Delta_0(U^2/16t)
[\ln(8t/U)-b_0]\ln(8t/T_c)$,
where $b_0=0.322$. 
For $U<U_c\sim 0.725\times 8t$,
a state with non-zero $\Delta_0$ is realized.
We have also applied the Ginzburg-Landau analysis to 
this mean field solution and found that, in the 2D case, the transition
temperature vanishes in accordance with the Mermin-Wagner-Coleman theorem.
Nevertheless, the above analysis demonstrates that in the ground state 
at zero temperature, the gap $\Delta_k$ is non-zero, 
and the system is in an insulating state.

We now further investigate the properties of the insulating phase.
In this phase, because the order parameter 
does not break the spin rotational symmetry,
there is no long-range magnetic order. 
However, a spin gap exists.
The spin-spin correlation function obtained
from the above mean field solution 
is ${\rm Im}\chi_s(q,\omega)/\omega 
\sim \langle 1/(2T\cosh^2 (\Delta_k/2T))
\rangle_{k\sim 0}$ for $\omega\rightarrow 0$. 
Here, $\langle\cdot\cdot\cdot\rangle_{k\sim 0}$ is
the angular average near $k=0$.
This spin gap behavior can be observed using NMR measurements.

Another important property of the insulating phase involves 
the charge degrees of freedom.
The formation of the gap $\Delta_k$ gives rise to a difference between
the charge densities at the sites 1 and 2 in a unit cell given by
$\rho_1-\rho_2\sim \Delta_0/t$, up to a constant factor.
Thus charge ordering (CO) with a charge density displacement
proportional to the gap characterizes this insulating state 
(see FIG.5(a)).
This noteworthy result can be understood as follows.
In our system, three electrons occupying nearest neighbor sites
cost energy loss caused by magnetic frustration.
Conversely, magnetic frustration induces 
an effective finite-range repulsion 
between electrons at nearest neighbor sites.
If this finite-range repulsion is sufficiently strong to
overcomes the on-site Coulomb interaction $U$,
the CO state will be stabilized. 
This is possible if $U$ is not so large. 
As $U$ increases, a transition to a conventional Mott insulator 
with no charge ordering should occur. 
This transition cannot be described within our weak-coupling analysis.  

\begin{figure}[h]
\centerline{\epsfxsize=7.8cm \epsfbox{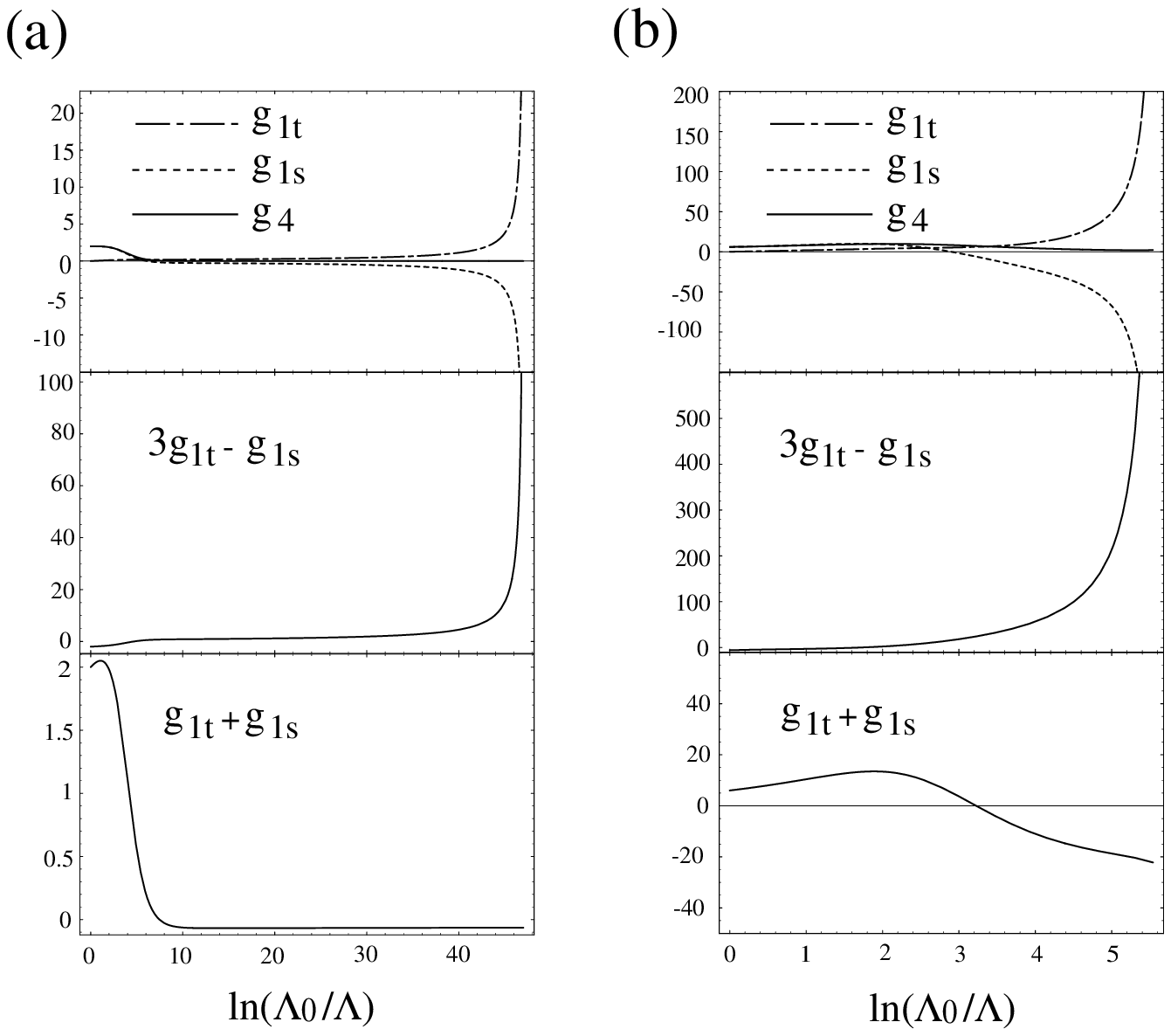}}
{FIG.3(a) The RG flow of the running couplings in the 2D case with 
$U/8t=0.25$.
(b) The RG flow in the 3D case with $U/8t=0.75$.}
\end{figure}

\begin{figure}[h]
\centerline{\epsfxsize=7.1cm \epsfbox{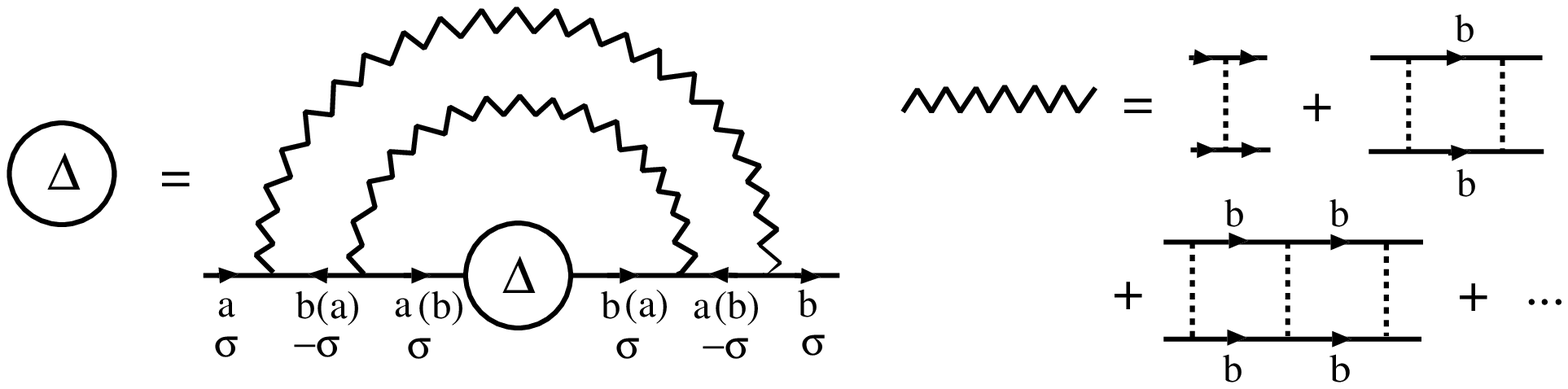}}
{FIG.4 Diagrams for the linearized gap equations.}
\end{figure}

{\it 3D pyrochlore--}
The above analysis can be straightforwardly 
generalized to the case of a 3D pyrochlore lattice.
We obtain the RG flow numerically from
(\ref{rge1})-(\ref{rge4}) for $d=3$.
Here, in contrast to the 2D case, for sufficiently small $U$
all couplings are irrelevant, and thus the semi-metal state is stable.
However, for values of $U$ larger than a certain critical value 
but still smaller than the band width,
RG flow similar to that in the 2D case is obtained, as shown in FIG.3(b).
The coupling $3g_{1t}-g_{1s}$, which is related to
the charge degrees of freedom, scales into the strong-coupling regime.
This RG flow implies that, as in the 2D case, 
a particle-hole pairing state with order parameters 
$\Delta^{(13)}_k=\sum_{\sigma}
\langle a^{\dagger}_{k1\sigma}a_{k3\sigma}\rangle$ and  
$\Delta^{(23)}_k=\sum_{\sigma}
\langle a^{\dagger}_{k2\sigma}a_{k3\sigma}\rangle$ is realized.
Although the value of $U$ used here is relatively large, 
we expect that the one-loop RG calculation still gives qualitatively
correct results, as long as $U$ is smaller than the band width.
To examine the validity of the one-loop calculation, 
we explore the self-consistent mean field solution.  
The self-consistent gap equations for the 3D case are
also given by the diagram shown in FIG.4, from which we find that
the gap functions are given by  
$\Delta^{(13)}_k=\sum_{\nu=1}^4 s_{\nu 1}(k)s_{\nu 3}(k)\Delta_{\nu}^{(13)}$,
$\Delta^{(23)}_k=\sum_{\nu=1}^4 s_{\nu 2}(k)s_{\nu 3}(k)\Delta_{\nu}^{(23)}$.
Using the symmetry properties of $s_{\mu\nu}(k)$ in momentum space,
we can show without solving the gap equations that $\Delta_{\nu}^{(13)}=0$,
and $\Delta_{1}^{(23)}=\Delta_{2}^{(23)}=\Delta_{3}^{(23)}$. Thus, from 
the orthogonal relations among the $s_{\mu\nu}(k)$, we have 
$\Delta^{(23)}_k=s_{42}(k)s_{43}(k)(\Delta_4^{(23)}-\Delta_1^{(23)})$. 
The quantity $\Delta_4^{(23)}-\Delta_1^{(23)}$ 
is determined from the gap equation.
According to the RG analysis, the transition occurs only for sufficiently
large $U$. Therefore, to determine the transition temperature and 
the gap function correctly,
we need to take into account 
the self-energy corrections i.e., pair breaking effect.
This calculation is rather involved, and we have not yet carried it out.
However, we see from the RG flow that at the critical temperature 
$T_c\sim \Lambda_0 e^{-l_c}=8t\times 0.004$, 
a transition from a semi-metal to an insulator occurs.
In the resulting insulating state, the three-fold degeneracy at
the $\Gamma$ point in the semi-metal state is lifted completely, and
a spin gap as well as a charge gap exists.

\begin{figure}[h]
\centerline{\epsfxsize=4.5cm \epsfbox{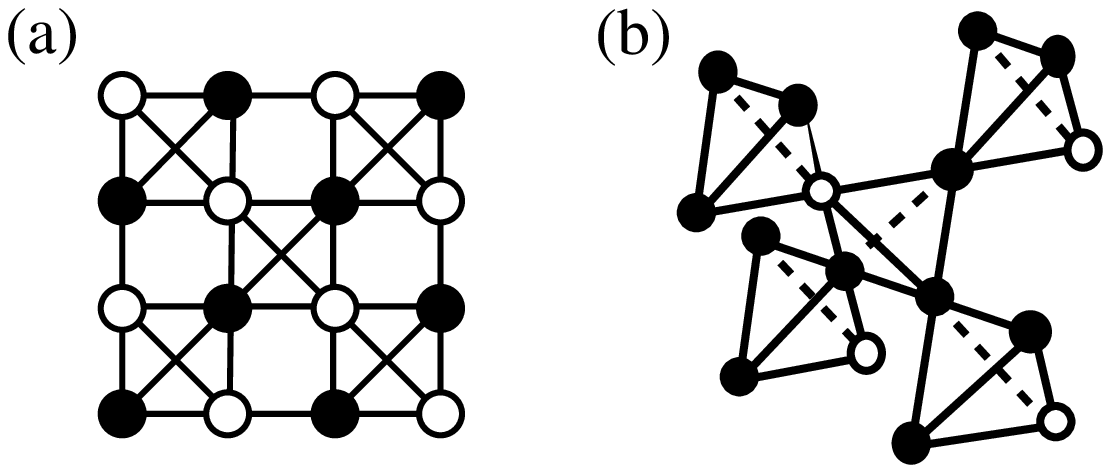}}
{FIG.5 (a) The CO pattern in the 2D case. (b) The CO pattern in
the 3D case.}
\end{figure}

As in the 2D case, we examine now the possibility of a CO state
in the 3D system. In this case, there are four sites in a unit cell.
The appearance of a gap causes a charge density displacement
on each site given by
$\delta\rho_{\nu}=2\sum_ks_{\nu 2}(k)s_{\nu 3}(k)\Delta^{(23)}_k$
for $\nu=1,2,3$. 
Using the symmetry properties of $s_{\mu\nu}(k)$, we find
$\delta\rho_1=\delta\rho_2=\delta\rho_3\neq 0$ 
and $\delta\rho_4=-3\delta\rho_1$. It is
thus found that CO with the pattern displayed in FIG.5(b) 
occurs in the insulating phase.
Interestingly, a similar CO pattern is observed in
the spinel system ${\rm AlV_2O_4}$ which possesses a V-site 
corner-sharing tetrahedron network\cite{al}.

We now apply the results obtained above 
to the description of the MIT observed in ${\rm Tl_2Ru_2O_7}$.
The band calculation gives the band width of this system 
$8t \sim 2$eV\cite{ogu}.
Experimental data on the size of $U$ do not exists.
However, typically, the value of $U$ for transition metal oxides
is $\sim 2$eV. 
This gives us reason to believe that the analysis given in this paper,
which suggests that the MIT occurs for
large $U\sim 8t$, can be applied to the description of the 
${\rm Tl_2Ru_2O_7}$ system.
The transition temperature estimated from the RG analysis is
$T_c\sim 96$K, which is almost comparable with 
the experimental values $100\sim 120$K\cite{take}.
A recent NMR measurement has revealed the presence of a spin gap
in the insulating state, which is consistent with our results\cite{tl}.
The possible existence of a CO state and 
large enhancement of charge fluctuations above $T_c$
predicted in our theory have not yet been investigated
experimentally. 
The experimental determination of whether a CO state exists
for ${\rm Tl_2Ru_2O_7}$ is a crucial test of this theory.
When there exists coupling to a lattice, 
CO should accompany lattice distortion.
It has been found experimentally that, in ${\rm Tl_2Ru_2O_7}$,
the lattice structure changes
from cubic to orthorhombic at the MIT point.
This observation seems to suggest the presence of 
large charge fluctuations in this system.

In conclusion, the 2D and 3D pyrochlore Hubbard models at the half-filling
show the transition from semi-metal to spin-gapped insulator.
In the insulating state, charge ordering occurs so as to relax 
geometrical frustration.
 
The author is grateful to K. Yamada and H. Tsunetsugu
for comments and discussions.
This work was supported by a Grant-in-Aid from the Ministry
of Education, Science, and Culture, Japan.
   

\end{multicols}
                                                                    
\end{document}